\pdfoutput=1
\RequirePackage{ifpdf}
\ifpdf 
\documentclass[pdftex]{sigma}
\else
\documentclass{sigma}
\fi

\numberwithin{equation}{section}

\begin{document}


\renewcommand{\thefootnote}{$\star$}

\newcommand{\arXivNumber}{1505.01588}

\renewcommand{\PaperNumber}{085}

\FirstPageHeading

\ShortArticleName{On Integrable Perturbations of Some Nonholonomic Systems}

\ArticleName{On Integrable Perturbations\\ of Some Nonholonomic Systems\footnote{This paper is a~contribution to the Special Issue
on Analytical Mechanics and Dif\/ferential Geometry in honour of Sergio Benenti.
The full collection is available at \href{http://www.emis.de/journals/SIGMA/Benenti.html}{http://www.emis.de/journals/SIGMA/Benenti.html}}}

\Author{Andrey V.~{TSIGANOV}~$^{\dag\ddag}$}

\AuthorNameForHeading{A.V.~Tsiganov}

\Address{$^\dag$~St.~Petersburg State University, St.~Petersburg, Russia}
\EmailD{\href{mailto:andrey.tsiganov@gmail.com}{andrey.tsiganov@gmail.com}}

\Address{$^\dag$~Udmurt State University, 1~Universitetskaya Str., Izhevsk, Russia}

\ArticleDates{Received May 08, 2015, in f\/inal form October 16, 2015; Published online October 20, 2015}

\Abstract{Integrable perturbations of the nonholonomic Suslov, Veselova, Chaplygin and Heisenberg problems are discussed in the framework of the classical Bertrand--Darboux method. We study the relations between the Bertrand--Darboux type equations, well stu\-died in the holonomic case, with their nonholonomic counterparts and apply the results to the construction of nonholonomic integrable potentials from the known potentials in the holonomic case.}


\Classification{37J60; 70G45; 70H45}

\rightline{\em Dedicated to Sergio Benenti on the occasion of his 70th birthday}

\renewcommand{\thefootnote}{\arabic{footnote}}
\setcounter{footnote}{0}

\section{Introduction}

In classical mechanics, the Euler--Poisson equations
\begin{gather}\label{eq-eul-jac}
{\mathbf I}\dot{\omega}={\mathbf I}\omega\times \omega+\gamma\times \frac{\partial V(\gamma)}{\partial\gamma},\qquad \dot{\gamma}=\gamma\times \omega
\end{gather}
describe the rotation of a rigid body with a f\/ixed point using a rotating reference frame with its axes f\/ixed in the body and parallel to the body's principal axes of inertia. Here $\omega=(\omega_1,\omega_2,\omega_3)$ is the angular velocity vector of the body, ${\mathbf I}=\mathrm{diag}(I_1,I_2,I_3)$ is a tensor of inertia, $\gamma=(\gamma_1,\gamma_2,\gamma_3)$ is a unit Poisson vector and~$V(\gamma)$ is a potential f\/ield. All the vectors are expressed in the so-called body frame and
 $x\times y$ means the cross product of two vectors in three-dimensional Euclidean space.

Let us impose a nonholonomic constraint on the angular velocity
\begin{gather*}
f=(\omega,a)=0\qquad\mbox{or}\qquad f=(\omega,\gamma)=0,
\end{gather*}
where $a$ is a f\/ixed unit vector in the rotating frame for the Suslov problem~\cite{sus46},
 $\gamma$ is a f\/ixed unit vector in the stationary frame for the Veselova problem~\cite{ves86} and $(x,y)$ means the scalar product of two vectors.
 In this case the Euler--Poisson equations~(\ref{eq-eul-jac}) are replaced by equations
\begin{gather}\label{eq-poi-sus}
{\mathbf I} \dot{\omega}={\mathbf I} \times \omega+\gamma\times \frac{\partial V(\gamma)}{\partial\gamma}+\lambda n,\qquad \dot{\gamma}=\gamma\times \omega,\qquad n=a,\gamma,
\end{gather}
 where $\lambda$ is a Lagrange multiplier which has to be found from the condition $\dot{f}=0$.

Both systems of dif\/ferential equations (\ref{eq-eul-jac}) and (\ref{eq-poi-sus})
are geometrically interpreted in terms of the vector f\/ield~$X$
\begin{gather}\label{vf}
\dot{x}_i = X_i(x_1,\dots, x_6)
\end{gather}
in a six-dimensional manifold $M$ with coordinates $ x=(\omega,\gamma)$.
 The classical Euler--Jacobi theorem says that the vector f\/ield~$X$~(\ref{vf}) on a six-dimensional manifold~$M$ is integrable by quadratures if it has an invariant volume form (invariant measure) and four functionally independent f\/irst integrals~\cite{koz13,ts13}.

 Equations~(\ref{eq-eul-jac}) and (\ref{eq-poi-sus}) preserve the norm of the unit Poisson vector $\gamma$
\begin{gather*}
C_1=(\gamma,\gamma)=\gamma_1^2+\gamma_2^2+\gamma_3^2=1,
\end{gather*}
and mechanical energy
\begin{gather*}
H_1=\frac{1}{2} (I\omega, \omega)+\frac{1}{2} V(\gamma).
\end{gather*}
An additive perturbations (\ref{eq-poi-sus}) of the Euler--Poisson vector f\/ield (\ref{eq-eul-jac}) change the standard invariant volume form and the second geometric f\/irst integral
\begin{gather*}
C_2=(\omega,{\mathbf I}\gamma) \to C_2=(\omega,n),
\end{gather*}
see details in~\cite{bm14}. Thus, according to the Euler--Jacobi theorem equations~(\ref{eq-eul-jac}) and~(\ref{eq-poi-sus}) are integrable by quadratures if there is one more independent f\/irst integral~$H_2$.

 There are several methods to uncover integrals of motion. A~one of the simplest method considers function~$H_2$ polynomial in the velocities, with the coef\/f\/icients being arbitrary functions of the coordinates, and requires that it is conserved in time
 \begin{gather}\label{m-eqint}
\dot{H_2}=0.
\end{gather}
 This condition yields a system of coupled partial dif\/ferential equations on the coef\/f\/icients, which are some of the most well-studied f\/irst-order PDE's in classical mechanics~\cite{koz-book}.

In this paper, we want to study what is going on with these well-known PDE's when we impose nonholonomic constraints on the rigid body motion. Some partial solutions of these new PDE's are discussed in literature, see, e.g., \cite{bm08,dr98,libr} and references within. Our main aim is to prove that these PDE's for the nonholonomic Chaplygin, Suslov and Veselova systems can be easily reduced to the well-studied PDE's for the Hamiltonian vector f\/ield~(\ref{eq-eul-jac}). Consequently, we can directly obtain all the possible integrable perturbations of these nonholonomic systems directly from the well-known integrable potentials of the Hamiltonian mechanics.

 The necessary references to the main aspects of nonholonomic mechanics can be found in several papers of Sergio Benenti dedicated to the analysis
of nonholonomic mechanical systems \cite{ben96,ben07,ben08,ben11}. Following in the steps of these papers we will consider
nonholonomic systems using only the knowledge of the basic notions of analytical mechanics, i.e., utilizing
a `user-friendly' approach to the dynamics of nonholonomic systems proposed by Sergio Benenti~\cite{ben07}.

 This paper is organized as follows. In Section~\ref{section2} we will introduce Bertarnd--Darboux equation for the holonomic particle on the plane and its counterpart for the holonomic particle on the sphere. Section~\ref{section3} contains the main result of integrable perturbations for the Suslov system. We will show that integrable potentials for the nonholonomic Suslov problem satisfy to the standard Bertarnd--Darboux equation for the holonomic particle in the plane. Sections~\ref{section4} and~\ref{section5} are devoted to the review of the known integrable potentials for the Veselova and Chaplygin systems. We will give the Bertrand--Darboux type equations for these systems and will show how these equations are reduced to the Bertarnd--Darboux equation for the holonomic particle on the sphere.
Section~\ref{section6} deals with two nonholonomic systems on the plane and contains the Bertrand--Darboux type equations for the nonholonomic oscillator and the Heisenberg system (nonholonomic integrator). Finally, we brief\/ly discuss the quasi-integrable potentials for the Suslov and Veselova problems, which were introduced by Llibre et al.

\section{Integrable potentials on the plane and sphere}\label{section2}

In \cite{bert} Bertrand studied the Newton equations for a particle on the plane
 \begin{gather}\label{newton-eq}
 \frac{d^2 q_1}{dt^2}=F_1,\qquad \frac{d^2q_2}{dt^2}=F_2,\qquad F_k=-\frac{\partial V(q_1,q_2)}{\partial q_k}
 \end{gather}
 and tried to solve equation (\ref{m-eqint}) using linear, quadratic or fractional (linear/linear) anzats in the velocity for the additional integral of motion.

In particular, according to Bertrand~\cite{bert}, if
 \begin{gather*}
 H_2=\sum_{i,j=1}^2 K_{ij}(q_1,q_2)\dot{q}_i\dot{q}_j + U(q_1,q_2),
 \end{gather*}
 then one equation $\dot{H}_2=0$ yields two systems of PDE's. The generic solution of the f\/irst system of equations for the coef\/f\/icients $K_ {ij} (q_1, q_2) $
\begin{gather*}
H_2= \left(-\frac{\alpha}{2}q_2^2-\beta_2q_2+\frac{\gamma_{11}}{2}\right)p_1^2
 +\left(-\frac{\alpha}{2}q_1^2-\beta_1q_1+\frac{\gamma_{22}}{2}\right)p_2 \\
\hphantom{H_2=}{}
 + (\alpha q_1q_2+\beta_1q_2+\beta_2q_1+\gamma_{12} )p_1p_2+U(q)
\end{gather*}
depends on the six constants of integration $\alpha$, $\beta_{1}$, $\beta_2$, $\gamma_{11}$, $\gamma_{12}$, $\gamma_{22}$.

 In order to describe the forces $F_{1,2}$ that should act on the particle Bertrand also extracted one linear second-order PDE on potential $V$ from the coupled system of equations on the poten\-tials~$V$ and~$U$
 \begin{gather}
(\alpha
q_1q_2+\beta_1q_1+\beta_2q_2+\gamma_{12})(\partial_{22}V-\partial_{11}V)\nonumber\\
\qquad{}
+\big(\alpha q_1^2-\alpha
q_2^2+2\beta_1q_1-2\beta_2q_2+\gamma_{11}-\gamma_{22}\big)\partial_{12}V\nonumber\\
\qquad{}
+3(\alpha q_1+\beta_1)\partial_2V-3(\alpha
q_2+\beta_2)\partial_1V=0, \label{bd-eq}
 \end{gather}
where $\partial_i={\partial}/{\partial q_i}$ and $ \partial_{ik} ={\partial^2}/{\partial q_i\partial q_k}$.

In \cite{bert} Bertand studied only some partial solutions of this equation, whereas in \cite{darb}
Darboux gave a complete solution and, therefore, now equation (\ref{bd-eq}) is called the Bertrand--Darboux equation. Later on Darboux results were included almost verbatim in the classical text of Whittaker on analytical mechanics, see historical details in \cite{sm08}.

Ideas used by Bertarnd and Darboux to solve the Bertrand--Darboux problem were genera\-li\-zed to study Hamiltonian systems def\/ined in Euclidean spaces of higher dimensions and in other (pseudo) Riemannian manifolds.
 According to Eisenhard~\cite{eis34}, the f\/irst system of PDE's is the Killing equation for the Killing tensor of second order with vanishing Haantjes torsion, hereafter called the characteristic Killing tensor on the Riemannian manifold~$Q$. The second system of PDE's on the various Riemannian manifolds was studied by Kalnins, Miller~\cite{km86}, Benenti~\cite{ben93, ben87} etc.

 In particular, Benenti formulated the following proposition.

 \begin{proposition}
 A natural Hamiltonian $H = \sum {\mathrm g}_{ij}p_ip_j + V$ on the cotangent bundle $T^*Q$ of a~Riemannian
manifold $Q$ is separable in orthogonal coordinates iff on~$Q$ there exists a~Killing~$K$ of second order with simple
eigenvalues and normal eigenvectors, so that
\begin{gather}\label{char-eq}
{\mathrm d}(K{\mathrm d}V)=0.
\end{gather}
 \end{proposition}

 Separable Hamiltonian f\/low has a necessary number of f\/irst integrals, which can be directly calculated from the characteristic Killing tensor $K$, which satisf\/ies to the Killing equation
\begin{gather*}
\nabla_\alpha K_{\beta\gamma}+\nabla_\beta K_{\gamma\alpha}+\nabla_\gamma K_{\alpha\beta}=0,
\end{gather*}
 where $\nabla$ is the Levi-Civita connection of the Riemannian metric \cite{ben93, ben87}. Tensor $K$ has normal eigenvectors if and only if its Haantjes torsion is equal to zero. Integrable systems associated with the Killing tensors of second order with nontrivial Haantjes torsion were found only recently~\cite{ts15}.

\subsection[Bertrand-Darboux type equation on the sphere]{Bertrand--Darboux type equation on the sphere}

Let us consider the standard Hamiltonian vector f\/ield describing rotation of a rigid body f\/ixed at the point
\begin{gather}\label{eul-eq}
\dot{M}=M\times \omega,\qquad \dot{\gamma}=\gamma\times \omega.
\end{gather}
Here $M$ is the angular momentum, $\omega= {\mathbf A} M$ is the angular velocity, $\gamma$ is the constant unit vector in a moving frame an~$ {\mathbf A}$ is the diagonal
inverse to $\mathbf I$ matrix
\begin{gather}\label{a-mat}
 {\mathbf A}={\mathbf I}^{-1}=\left(
 \begin{matrix}
 {a}_1 & 0 & 0 \\
 0 & {a}_2 & 0 \\
 0 & 0 & {a}_3
 \end{matrix}
 \right),\qquad a_k=\frac{1}{I_k}.
\end{gather}
According to Euler, there are two f\/irst integrals of second order in momenta
\begin{gather*}
H_1=\frac{1}{2} (M,{\mathbf A} M),\qquad H_2=(M,M)
 \end{gather*}
and two geometric integrals
\begin{gather*}
C_1=\gamma_1^2+\gamma_2^2+\gamma_3^2,\qquad C_2=\gamma_1M_1+\gamma_2M_2+\gamma_3M_3,
\end{gather*}
which are the Casimir functions of the underlying Poisson structure.

Let us consider perturbation of the free motion (\ref{eul-eq}) by adding forces associated with potential f\/ield $V_1(\gamma)$. For the perturbed Hamiltonian vector f\/ield
\begin{gather}\label{eul-eqV}
\dot{M}=M\times \omega+\gamma\times \frac{\partial V_1(\gamma)}{\partial\gamma},\qquad \dot{\gamma}=\gamma\times \omega,
\end{gather}
functions $C_{1,2}$ are also constants of motion and, therefore, we can exclude the third component of the Poisson vector $\gamma$ from other calculations
\begin{gather*}
\gamma_3=\sqrt{1-\gamma_1^2-\gamma_2^2}.
\end{gather*}
Substituting the polynomials of second order in momenta
\begin{gather}
\label{eul-ham}
H_1=\frac{1}{2}(M,{\mathbf A} M)+V_1(\gamma),\qquad H_2=(M,M)+V_2(\gamma)
\end{gather}
in the equations $\dot{H}_{1,2}=0$ one gets three partial dif\/ferential equations on $V_{1,2}(\gamma)${\samepage
 \begin{gather}
\partial_2(a_1V_2-2V_1)=0,\qquad
\partial_1(a_2V_2-2V_1)=0,\nonumber\\
a_3 (\gamma_2\partial_1V_2-\gamma_1\partial_2V_2 )+
2\gamma_1\partial_2V_1-2\gamma_2\partial_1V_1=0.\label{gensys-eul}
\end{gather}
Here $V_{1,2}(\gamma)$ are functions on two independent variables $\gamma_1$, $\gamma_2$ and $\partial/\partial \gamma_k=\partial_k$.}

The generic solution of these equations (\ref{gensys-eul}){\samepage
\begin{gather*}
 V_1=-\frac{1}{2}\bigl(a_2a_3\gamma_1^2+a_1a_3\gamma_2^2+a_1a_2\gamma_3^2\bigr),\qquad V_2=a_1\gamma_1^2+a_2\gamma_2^2+a_3\gamma_3^2
 \end{gather*}
is associated with the Clebsh system \cite{koz-book}.}

At $C_2=(\gamma,M)=0$ the phase space is equivalent to the cotangent bundle of the two-dimensional sphere~\cite{bog85}. In this case equation $\dot{H}_{2}=0$ yields only two equations
 \begin{gather}
 \bigl(\gamma_2^2(a_2-a_3)-a_2\bigr)\partial_1V_2-\gamma_1\gamma_2(a_1-a_3)\partial_2V_2+2\partial_1 V_1=0,
 \nonumber \\
 \bigl(\gamma_1^2(a_1-a_3)-a_1\bigr)\partial_2V_2-\gamma_1\gamma_2(a_2-a_3)\partial_1V_2+2\partial_2 V_1 =0,\label{uv-geq}
 \end{gather}
 from which we can easily get the equation on one potential. Namely, when equations are dif\/ferentiated by
 $\gamma_{1,2}$ and subtracted from each other, the result is
 \begin{gather}
 \partial_1\big(\big(\gamma_2^2(a_2-a_3)-a_2\big)\partial_1V_2-\gamma_1\gamma_2(a_1-a_3)\partial_2V_2\big)
 \nonumber \\
 \qquad{}-\partial_2\big(\big(\gamma_1^2(a_1-a_3)-a_1\big)\partial_2V_2-\gamma_1\gamma_2(a_2-a_3)\partial_1V_2\big) =0.\label{uv-geq2}
 \end{gather}
 It is well-known that the characteristic equation (\ref{char-eq}), (\ref{uv-geq2})
 has a continuum of solutions labelled by two arbitrary functions $G_{1,2}$
\begin{gather}\label{uv-sol}
 V_1=\frac{u_2G_1(u_1)-u_1G_2(u_2)}{2(u_2-u_1)},\qquad
 V_2=\frac{G_1(u_1)-G_2(u_2)}{u_2-u_1}.
\end{gather}
Here $u_1$, $u_2$ are sphero-conical coordinates on the sphere
 \begin{gather*}
\gamma_i=\sqrt{\frac{(u_1-a_i)(u_2-a_i)}{(a_j-a_i)(a_m-a_i)}},\qquad i\neq j\neq m.
\end{gather*}
If $p_{u_{1,2}}$ are the corresponding momenta def\/ined by relations
\begin{gather*}
M_i=\frac{2\varepsilon_{ijm}\gamma_j\gamma_m(a_j-a_m)}{u_1-u_2}\big((a_i-u_1)p_{u_1}-(a_i-u_2)p_{u_2}\big),
\end{gather*}
then f\/irst integrals $H_{1,2}$~(\ref{eul-ham}) satisfy to the following separation relations
\begin{gather}\label{ell-sep-rel}
4(u_i-a_1)(u_i-a_2)(u_i-a_3)p_{u_i}^2+G_i(u_i)-u_iH_2+2H_1=0,\qquad i=1,2.
\end{gather}

If we take homogeneous polynomials of $N$-th order $G_{1,2}=u^N$, the corresponding poten\-tials~$V_{1,2}^{(N)}$ satisfy to the well-known recurrence relations
\begin{gather*}
V_1^{(1)}=0,\qquad V_2^{(1)}=1,\qquad 2V_1^{(N)}=\rho V_2^{(N-1)},\qquad V_2^{(N)}=\sigma V_2^{(N-1)}-2V_1^{(N-1)},
\end{gather*}
where
\begin{gather}
\sigma=(u_1+u_2)=(a_1+a_2+a_3)-a_1\gamma_1^2-a_2\gamma_2^2-a_3\gamma_3^2,\nonumber\\
\rho=u_1u_2=a_2a_3\gamma_1^2+a_1a_3\gamma_2^2+a_1a_2\gamma_3^2.\label{sigma-rho}
\end{gather}
According to Bogoyavlenskii \cite{bog85} these potentials are equal to
\begin{gather}
V_2^{(N)} = \sum_{k=0}^{[N/2]}(-1)^k\left(
 \begin{matrix}
 N-k \\
 k
 \end{matrix}
 \right) \rho^k\sigma^{N-2k},\nonumber\\
 2V_1^{(N)} = \sum_{k=0}^{[(N-1)/2]}(-1)^k\left(
 \begin{matrix}
 N-k -1\\
 k
 \end{matrix}
 \right) \rho^{k+1}\sigma^{N-2k-1}. \label{uv-nell}
\end{gather}
Here $[z]$ is an 	integer part of the rational number~$z$. If $G_{1,2}(u)=u^{-K}$, one gets rational potentials
\begin{gather*}
V_2^{(-K)}=\frac{1/u_1^K-1/u_2^K}{u_2-u_1}=\frac{1}{(u_1u_2)^K}\frac{u_2^K-u_1^K}{u_2-u_1}
=-\frac{V_2^{(K)}}{\rho^K}
\end{gather*}
and
\begin{gather*}
2V_1^{(-K)}=\frac{u_2/u_1^K-u_1/u_2^K}{u_2-u_1}=\frac{1}{(u_1u_2)^K}\frac{u_2^{K+1}-u_1^{K+1}}{u_2-u_1}
=-\frac{V_2^{(K+1)}}{\rho^K}
\end{gather*}
Of course, any linear combination of these polynomial and rational potentials also satisf\/ies~(\ref{uv-geq}).

For instance, at $N=2$ one gets the Neumann system
\begin{gather*}
 H_1^{(2)}=\frac{1}{2} (M,{\mathbf A} M)-a_2a_3\gamma_1^2-a_1a_3\gamma_2^2-a_1a_2\gamma_3^2,\\
 H_2^{(2)}=(M,M)+a_1\gamma_1^2+a_2\gamma_2^2+a_3\gamma_3^2,
\end{gather*}
 and the Braden system at $K=1$
\begin{gather*}
 H_1^{(-1)} = \frac{1}{2}
(M,{\mathbf A} M)+\frac{a_1\gamma_1^2+a_2\gamma_2^2+a_3\gamma_3^2-\big(\gamma_1^2+\gamma_2^2+\gamma_3^2\big)(a_1+a_2+a_3)}{a_2a_3\gamma_1^2+a_1a_3\gamma_2^2+a_1a_2\gamma_3^2}
,\\
H_2^{(-1)} = (M,M)+\frac{1}{a_2a_3\gamma_1^2+a_1a_3\gamma_2^2+a_1a_2\gamma_3^2}.
\end{gather*}
In \cite{stef85} Wojciechowski presented another family of potentials associated with other symmetric functions $G_ {1,2}$~(\ref{uv-sol}) on variables $u_{1,2}$ and parameters $a_1$, $a_2$, $a_3$. For instance, if
\begin{gather*}
F^{(2)}_{1,2}(u)=u\bigl(u^2-(a_1+a_2+a_3)u+a_1a_2+a_1a_3+a_2a_3\bigr),\\
F^{(3)}_{1,2}(u)=u(u-a_1)(u-a_2)(u-a_3)
\end{gather*}
then the second integrals of motion read as
\begin{gather*}
H_2^{(2)}=(M,M)+\sum a_i^2\gamma_i^2-\left(\sum a_i\gamma_i^2\right)^2,\\
H_2^{(3)}=(M,M)+\sum a_i^3\gamma_i^2-2\left(\sum a_i\gamma_i^2\right)\left(\sum a_k^2\gamma_k^2\right)+\left(\sum a_i\gamma_i^2\right)^3.
\end{gather*}
 More complicated functions $G_{1,2}$ (\ref{uv-sol}) yield more complicated potentials, for example, rational functions
\begin{gather*}
G_{1,2}(u)=\sum_{i=1}^3 b_ib_j u+\frac{(a_1-a_3)(a_1-a_2)b_1^2}{u-a_1}
+\frac{(a_2-a_3)(a_2-a_1)b_2^2}{u-a_2}+\frac{(a_3-a_1)(a_3-a_2)b_3^2}{u-a_3}
\end{gather*}
give rise to the Rosochatius potentials \cite{ros77}
\begin{gather*}
H_1^{\rm (Ros)}=\frac{1}{2}
(M,{\mathbf A} M)+\frac{b_1^2\big(a_2\gamma_3^2+a_3\gamma_2^2\big)}{\gamma_1^2}+\frac{b_2^2\big(a_1\gamma_3^2+a_3\gamma_1^2\big)}{\gamma_2^2}
+\frac{b_3^2(a_1\gamma_2^2+a_2\gamma_1^2)}{\gamma_3^2},\\
H_2^{\rm (Ros)}=(M,M)+\frac{b_1^2}{\gamma_1^2}+\frac{b_2^2}{\gamma_2^2}+\frac{b_3^2}{\gamma_3^2}
\end{gather*}
up to the constant terms, see also~\cite{bj}.

 In \cite{dr02} Dragovi\'c considered functions $G_{1,2}$ (\ref{uv-sol}), which are the Laurent polynomials, and ingeniously coupled the corresponding potentials~$V_{1,2}$ together with
the Appell hypergeometric function.

These 	familiar well-studied potentials are	permanently	rediscovered both in holonomic~\cite{bj,val07} and nonholonomic mechanics \cite{fed04,libr}.

\section{Suslov problem}\label{section3}

One of the most widely known mechanical nonholonomic systems is the Suslov problem descri\-bing motion of a rigid body under the following constraint on its angular velocity
\begin{gather}\label{rel-sus}
(\omega,a)=0,
\end{gather}
where $a$ is a f\/ixed unit vector in the body frame \cite{sus46}. It means that there is no twisting around this vector $a$.

Imposing this constraint we have to add some terms with the Lagrangian multiplier to the initial Hamiltonian vector f\/ield
\begin{gather}\label{eq-sus}
\mathbf I \dot{\omega}=\mathbf I \times \omega+\gamma\times \frac{\partial V_1(\gamma)}{\partial\gamma}+\lambda a,\qquad \dot{\gamma}=\gamma\times \omega.
\end{gather}
Dif\/ferentiating the constraint (\ref{rel-sus}) by time and using the equation of motion we obtain
\begin{gather*}
\lambda=\frac{1}{\big(\mathbf I^{-1}a,a\big)}
\left(
{\mathbf I}^{-1}a,{\mathbf I} \times \omega+\frac{1}{2} \gamma\times \frac{\partial V_1(\gamma)}{\partial\gamma}
\right).
\end{gather*}
Vector f\/ield (\ref{eq-sus}) preserves the mechanical energy
\begin{gather}\label{sus-energy}
 {H}_1=(M,{\mathbf A}M)+V_1(\gamma)
 \end{gather}
and the geometric constants of motion
\begin{gather*}
C_1=(\gamma,\gamma)=1,\qquad C_2=(\omega,a)=0,
\end{gather*}
which allows us to remove the redundant variable from the calculations
\begin{gather*}
\gamma_3=\sqrt{1-\gamma_1^2-\gamma_2^2}.
\end{gather*}
If we assume that $a$ is an eigenvector of the tensor of inertia \cite{bm08,dr98,libr}, i.e., that tensor of inertia is diagonal
\begin{gather*}
\mathbf I=\left(
 \begin{matrix}
 I_1 & 0 & 0 \\
 0 & I_2 & 0 \\
 0 & 0 & I_3
 \end{matrix}
 \right),
\end{gather*}
 and vector $a$ is equal to $a=(0,0,1)$ in some coordinate frame, the constraint is trivial
 \begin{gather*}
 \omega_3=0.
 \end{gather*}
 Substituting the standard anzats for the second integral of motion
 \begin{gather*}
 H_2=f_1(\gamma)\omega_1^2+f_2(\gamma)\omega_2^2+f_3(\gamma)\omega_1\omega_2+V_2(\gamma)
 \end{gather*}
with unknown functions $f_k$ and $V_2$ on the components $\gamma_{1,2}$ of the Poisson vector in the equation $\dot{H}_2=0$ we obtain
 the well-known Bertrand--Darboux equation on potential~\cite{bert,darb}.

\begin{proposition}
For vector field \eqref{eq-sus} the following statements are equivalent:
\begin{enumerate}\itemsep=0pt
 \item[$1.$] There is an additional independent first integral of second order in velocities
 \begin{gather*}
 H_2=\left(\frac{\alpha\sqrt{I_1I_2}}{2}\gamma_1^2+\beta_1\sqrt{I_1}\gamma_1+\frac{\gamma_{11}\sqrt{I_1}}{2\sqrt{I_2}}\right)\omega_1^2\nonumber\\
\hphantom{H_2=}{}
 +\left(\frac{\alpha\sqrt{I_1I_2}}{2}\gamma_2^2+\beta_2\sqrt{I_2}\gamma_2+\frac{\gamma_{22}\sqrt{I_2}}{2\sqrt{I_1}}\right)\omega_2^2\nonumber\\
\hphantom{H_2=}{}
+\left(\alpha\sqrt{I_1I_2}\gamma_1\gamma_2+\beta_1\sqrt{I_1}\gamma_2+\beta_2\sqrt{I_2}\gamma_1+\gamma_{12}\right)\omega_1\omega_2
+V_2(\gamma).
 \end{gather*}
 \item[$2.$] Potential $V_1$ satisfies the Bertrand--Darboux equation \eqref{bd-eq}
with
 \begin{gather} \label{q-var-sus}
 q_1={\gamma_1}{\sqrt{I_2}},\qquad q_2={\gamma_2}{\sqrt{I_1}}.
 \end{gather}
\item[$3.$]
Potential~$V_1$ is separable. A characteristic coordinate system for the Bertrand--Darboux equation
provides separation for~$V_1$ and can be taken as one of the following four
orthogonal coordinate systems on the~$q_{1,2}$-plane: elliptic, parabolic, polar or Cartesian.
\end{enumerate}
\end{proposition}

The proof is completely similar to the one for the original Bertrand--Darboux theorem~\cite{darb,sm08}.

This result allows us to suppose that the nonholonomic Suslov system is equivalent to the holonomic motion on the plane with coordinates $q_1$, $q_2$ after some singular change of time, but its study is out of the framework of the present note.

Another reduction to the Bertrand--Darboux equation was proposed in \cite{libr}. Integrable vector f\/ield remains integrable for any f\/ixed value of mechanical energy $H_1$, for instance on the zero-energy hypersurface
\begin{gather*}H_1=0.\end{gather*}
In the Hamiltonian case the separation of variables of this null Hamilton--Jacobi equation is equivalent to the ordinary separation of the image of the original Hamiltonian under a generalized Jacobi--Maupertuis transformation \cite{ben05}.

If we substitute potential
\begin{gather*}
V_1=-\frac{1}{2}\left( \frac{mu_1^2}{I_2}+\frac{mu_2^2}{I_1}\right)
\end{gather*}
and velocities
\begin{gather*}
\omega_1=\frac{mu_2}{I_1} ,\qquad \omega_2=-\frac{\mu_1}{I_2},\qquad \omega_3=0
\end{gather*}
from Theorem 1 in \cite{libr} into the f\/irst integral $H_1$ (\ref{sus-energy}), one gets $H_1=0$. Here $\mu_{1,2}$ are functions on the components of the Poisson vector $\gamma_{1,2}$, which satisfy equations (\ref{eq-sus})
\begin{gather*}
\dot{\gamma_1}=-\sqrt{1-\gamma_1^2-\gamma_2^2} \frac{\mu_1}{I_2},\qquad \dot{\gamma_2}=-\sqrt{1-\gamma_1^2-\gamma_2^2} \frac{\mu_2}{I_1}.
\end{gather*}
Dif\/ferentiating these equations with respect to time we obtain the Newton equations (\ref{newton-eq}) on~$q_ {1,2}$~(\ref{q-var-sus}) with forces labelled by two functions $\mu_{1,2}(q_1,q_2)$. Thus, on the zero-energy hypersurface of the initial Hamiltonian one gets an initial Bertrand problem with the well-known solution. Of course, on this zero-energy hypersurface we can f\/ind other solutions associated with nonitegrable potentials on the whole phase space. In~\cite{libr} such potentials were called quasi-implicitly integrable or locally integrable potentials.

\section{Veselova system}\label{section4}

Let us consider the nonholonomic Veselova system describing the motion of a rigid body under the following constraint
\begin{gather}\label{rel-ves}
(\omega,\gamma)=0,
\end{gather}
where $\gamma$ is a unit Poisson vector f\/ixed in space \cite{ves86}. It means that there is no twisting around vector $\gamma$.

According to \cite{ves86} this constraint shifts the initial Hamiltonian vector f\/ield (\ref{eul-eq})
\begin{gather}\label{ves-eqm}
\dot{M}=M\times \omega +\lambda \gamma,\qquad \dot{\gamma}=\gamma\times \omega,
\end{gather}
where the Lagrangian multiplier $\lambda$ is chosen so that the constraint (\ref{rel-ves}) is satisf\/ied at any time
\begin{gather*}
\lambda=\frac{( {\mathbf A} M\times M, {\mathbf A} \gamma)}{( {\mathbf A} \gamma,\gamma)}.
\end{gather*}
There are integrals of motion of second order in momenta
\begin{gather*}
 {H}_1=\frac{1}{2} (M,{\mathbf A}M),\qquad {H}_2=(M,M)-(\gamma,\gamma)^{-1}(\gamma,M)^2
 \end{gather*}
and two geometric constants of motion
\begin{gather*}C_1=(\gamma,\gamma),\qquad C_2=(\gamma,\omega)=0.
\end{gather*}
In the presence of the potential f\/ield equations of motion (\ref{ves-eqm}) become
\begin{gather}\label{ves-eqV}
\dot{M}=M\times \omega+\lambda \gamma+ \gamma\times \frac{\partial W_1(\gamma)}{\partial\gamma},\qquad \dot{\gamma}=\gamma\times \omega,
\end{gather}
where
\begin{gather*}
\lambda=\frac{( {\mathbf A} M\times M+\gamma\times \partial W_1(\gamma)/\partial\gamma, {\mathbf A} \gamma)}{( {\mathbf A} \gamma,\gamma)}.
\end{gather*}
As usual, functions $C_{1,2}$ remain constants of motion, and we can exclude the redundant variable
\begin{gather*}
\gamma_3=\sqrt{1-\gamma_1^2-\gamma_2^2}.
\end{gather*}
Vector f\/ield (\ref{ves-eqV}) is a conformally Hamiltonian f\/ield, see, for instance, \cite{bm14, ts12}. Substituting
the following anzats for integrals of motion
\begin{gather}
\label{ves-ham}
 H_1=\frac{1}{2} (M,\mathbf A M)+W_1(\gamma),\qquad H_2=(M,M)-(\gamma,\gamma)^{-1}(\gamma,M)^2+W_2(\gamma)
\end{gather}
 in $\dot{H}_{1,2} =0$ one gets two f\/irst-order equations on potentials $W_{1,2}$
 \begin{gather}
 2\bigl(\gamma_2^2\big(a_2^{-1}-a_3^{-1}\big)-a_2^{-1}\bigr)\partial_1W_1-2\gamma_1\gamma_2\big(a_1^{-1}-a_3^{-1}\big)\partial_2W_1+\partial_1 W_2=0,
 \nonumber\\
 2\bigl(\gamma_1^2\big(a_1^{-1}-a_3^{-1}\big)-a_1^{-1}\bigr)\partial_2W_1-2\gamma_1\gamma_2\big(a_2^{-1}-a_3^{-1}\big)\partial_1W_1+\partial_2 W_2 =0.\label{uv-ves}
 \end{gather}
 \begin{proposition}
 After the inversion of parameters $a_k\to a_k^ {-1} $ and substitution
\begin{gather*}
W_2=2V_1,\qquad W_1=\frac{V_2}{2}
\end{gather*}
equations \eqref{uv-ves} coincide with equations \eqref{uv-geq} for potentials on the two-dimensional sphere.
\end{proposition}

Thus, all the integrable potentials for the nonholomic Veselova system are easily expressed via well-known integrable potentials~$V_ {1,2} $ for the holonomic system on the two-dimensional sphere. For functions~$G_{1,2}$~(\ref{uv-sol}) which are the Laurent polynomials in~$u_{1,2}$ these expressions were found in~\cite{dr98}. We only want to note that it is true for any integrable potentials.

Following Theorem~2 in~\cite{libr} let us f\/ix the values of velocities by equations
\begin{gather*}
I_1\omega_1\gamma_2-I_2\omega_2\gamma_1-\Psi_2=0,\qquad
p\omega_3-\Psi_1=0,\qquad \omega_1\gamma_1+\omega_2\gamma_2+\omega_3\gamma_3=0,
\end{gather*}
where $p=\sqrt{I_1I_2I_3\left(\frac{\gamma_1^2}{I_1}+\frac{\gamma_2^2}{I_2}+\frac{\gamma_3^2}{I_3}\right)}$ and $\Psi_{1,2}(\gamma)$ are functions on $\gamma$. Substituting these velocities and potential
\begin{gather*}
W_1=-\frac{\Psi_1^2+\Psi_2^2}{2(I_1\gamma_2^2+I_2\gamma_1^2)}
\end{gather*}
 into the mechanical energy $H_1$ (\ref{ves-ham}) one gets $H_1=0$.

 The remaining three equations of motion of the components of the Poisson vector $\gamma$ are easily reduced to equations of motion for the holonomic particle on the sphere with forces labelled by two functions $\Psi_{1,2}(\gamma)$. Thus, on the zero-energy hypersurface of the initial Hamiltonian one gets standard characteristic equation~(\ref{uv-geq2}) with the well-known solutions.

\section{Chaplygin ball}\label{section5}

As in \cite{ch03} we consider the rolling of a dynamically balanced ball on a horizontal absolutely rough table without slipping or sliding. `Dynamically balanced' means that the geometric center coincides with the center of mass, but mass distribution is not assumed to be homogeneous. Because of the roughness of the table this ball cannot slip, but it can turn about the vertical axis without violating the constraints.

After reduction \cite{ch03} motion of the Chaplygin ball is def\/ined by the following vector f\/ield
\begin{gather}\label{ch-eq}
\dot{M}=M\times \omega,\qquad \dot{\gamma}=\gamma\times \omega.
\end{gather}
Here $M$ is the angular momentum of the ball with respect to the contact point, $\omega$ is the angular velocity vector of the rolling ball. Its mass, inertia tensor and radius will be denoted by~$m$, ${\mathbf I} = \mathrm{diag}(I_1, I_2, I_3 )$ and~$b$ respectively. All the vectors are expressed in the so-called body frame, which is f\/irmly attached to the ball, and its axes coincide with the principal inertia axes of the ball.

The angular velocity vector is equal to $\omega={\mathbf A}_gM$, here matrix
\begin{gather*}
{\mathbf A}_g=\mathbf A +d\mathrm g(\gamma){\mathbf A} \gamma \otimes\gamma{\mathbf A}
\end{gather*}
is def\/ined by the nondegenerate matrix $\mathbf A$~(\ref{a-mat}) and function
\begin{gather}\label{fun-g}
{\mathrm g}(\gamma)=\frac{1}{1-d \big(a_1 \gamma_1^2+a_2 \gamma_2^2+a_3 \gamma_3^2\big)},\qquad d=mb^2.
\end{gather}
 It is easy to prove that vector f\/ield (\ref{ch-eq}) preserves two polynomials of second order in momenta
\begin{gather*}
H_1=\frac{1}{2} (M,{\mathbf A}_gM),\qquad H_2=(M,M)
\end{gather*}
and two geometric constants of motion
\begin{gather*}
C_1=\gamma_1^2+\gamma_2^2+\gamma_3^2=1,\qquad C_2=\gamma_1M_1+\gamma_2M_2+\gamma_3M_3,
\end{gather*}
see details in \cite{bm01,ts11,ts13}.

Indeed, equations of motion of the ball in the potential f\/ield
\begin{gather}\label{ch-eqV}
\dot{M}=M\times \omega+\gamma\times \frac{\partial U_1(\gamma)}{\partial\gamma},\qquad \dot{\gamma}=\gamma\times \omega
\end{gather}
 have the same form as the equations (\ref{eul-eqV}) in rigid body dynamics. In fact, the principal dif\/ference between holonomic and nonholonomic systems is hidden within the relation of the angular velocity to the angular momentum.

According to \cite{ts12b, ts12}, integrals of motion for the Veselova system are expressed via the integrals
of motion for the Chaplygin ball. So, we can easily express integrable potentials for the Chaplygin system via integrable potentials for the Veselova system $W_{1,2}$ (\ref{uv-ves}) and then via integrable potentials~$V_{1,2}$~(\ref{uv-geq}) for the holonomic system on the two-dimensional sphere. Of course, we can get the same result by directly substituting standard ansatz
\begin{gather}
\label{ch-ham}
 H_1=\frac{1}{2} (M,{\mathbf A}_gM)+U_1(\gamma),\qquad H_2=(M,M)+U_2(\gamma)
\end{gather}
in $\dot{H}_{1,2}=0$ we obtain
 \begin{gather*}
a_1(a_2-a_3)\gamma_1\gamma_2\partial_1U_2+a_1\bigl((a_2-a_3)\gamma_2^2+a_3-1\bigr)\partial_2U_2+2\mathrm g^{-1}\partial_2U_1=0,\\
a_2\bigl((a_1-a_3)\gamma_1^2+a_3-1\bigr)\partial_1U_2+a_2(a_1-a_3)\gamma_1\gamma_2\partial_2U_2-2\mathrm g^{-1}\partial_1U_1=0,\\
a_3\gamma_2\bigl((a_1-a_2)\gamma_1+a_2-1\bigr)\partial_1U_2+a_3\gamma_1\bigl((a_1-a2)\gamma_2^2-a_1+1\bigr)\partial_2U_2 \\
\qquad{}
+2{\mathrm g}^{-1}\bigl(\gamma_1\partial_2U_1-\gamma_2\partial_1U_1\bigr)=0.
 \end{gather*}
This system of equations has only one solution
 \begin{gather*}
 U_1=-\frac{1}{2} \big(a_2a_3\gamma_1^2+a_1a_3\gamma_2^2+a_1a_2\gamma_3^2\big),\qquad
 U_2=a_1\gamma_1^2+a_2\gamma_2^2+a_3\gamma_3^2,
 \end{gather*}
which coincides with the single solution of the initial system~(\ref{gensys-eul}) associated with the Clebsch model. This integrable potential has been found in~\cite{koz85}.

At $C_2=0$ conditions $\dot{H}_{1,2}=0$ are thus
\begin{gather}
\mathrm g \gamma_1 \gamma_2\bigl (a_2(a_1-a_3) \gamma_1^2+a_1(a_2-a_3) \gamma_2^2-a_1 a_2+(a_1+a_2-1) a_3\bigr)\partial_1U_2\nonumber\\
\qquad{}
-\mathrm g\bigl(a_1\big(x_1^2+x_2^2-1\big)\big(a_2x_2^2-1\big)-a_3\bigl(a_1 \big(x_2^4-1\big)+x_1^2\big(a_2x_2^2-1\big)\bigr)\bigr)\partial_2U_2
\nonumber\\
\qquad{}
-2\gamma_1 \gamma_2\partial_1U_1-2(\gamma_2^2-1)\partial_2U_1=0,
\nonumber\\
 \mathrm g\bigl(a_2\big(x_1^2+x_2^2-1\big)\big(a_1x_1^2-1\big)-a_3\bigl(a_2 \big(x_1^4-1\big)+x_2^2\big(a_1x_1^2-1\big)\bigr)\bigr)\partial_1U_2
 \nonumber\\
\qquad{} -\mathrm g \gamma_1 \gamma_2\bigl(a_2(a_1-a_3) \gamma_1^2+a_1(a_2-a_3) \gamma_2^2-a_1 a_2+(a_1+a_2-1) a_3\bigr)\partial_2U_2
\nonumber\\
\qquad{} +2(x_1^2-1)\partial_1U_1+ 2\gamma_1 \gamma_2\partial_2U_1=0.\label{uv-eqch}
\end{gather}
Here $\mathrm g\equiv\mathrm g(\gamma)$ is the function def\/ined by~(\ref{fun-g}). If we change the parameters
 \begin{gather*}
 e_1= \frac{a_1}{1-a_1},\qquad e_2= \frac{a_2}{1-a_2},\qquad
 e_3= \frac{a_3}{1-a_3}
 \end{gather*}
 and substitute in (\ref{uv-eqch})
\begin{gather}
2U_1 = \big(e_2e_3\gamma_1^2+e_1e_3\gamma_2^2+e_1e_2\gamma_3^2\big)V_2+2V_1,\nonumber\\
U_2=d\bigl(1+(e_1+e_2+e_3)-e_1\gamma_1^2-e_2\gamma_2^2-e_3\gamma_3^2+e_2e_3\gamma_1^2+e_1e_3\gamma_2^2+e_1e_2\gamma_3^2\bigr)V_2,\label{zam-ch}
\end{gather}
 then the equations (\ref{uv-eqch}) become
\begin{gather*}
 \bigl(\gamma_2^2(e_2-e_3)-e_2\bigr)\partial_1V_1-\gamma_1\gamma_2(e_1-e_3)\partial_2V_2+2\partial_1 V_1=0,
 \\
 \bigl(\gamma_1^2(e_1-e_3)-e_1\bigr)\partial_2V_1-\gamma_1\gamma_2(e_2-e_3)\partial_1V_2+2\partial_2 V_1 =0.
 \end{gather*}
 It is easy to see that this system coincides with the initial system of equations~(\ref{uv-geq}) def\/ining integrable potentials on the sphere up to $a_k\to e_k$.

 Thus, for the Chaplygin ball imposition of the nonholonomic constraint leads to deformation of potentials~(\ref{zam-ch}) and to replacement of parameters $a_k\to e_k$.

\begin{proposition}
At $C_2=0$ conformally Hamiltonian vector field \eqref{ch-eqV} has two integrals of motion \eqref{ch-ham} with potentials
\begin{gather*}
2U_1=\rho V_2+2V_1,\qquad U_2=d(\rho+\sigma+1)V_2.
\end{gather*}
Here $V_{1,2}$ are integrable potentials on the sphere~\eqref{uv-sol} after replacement of parameters $a_k\to e_k$, and
\begin{gather*}
\sigma = \big(\gamma_1^2+\gamma_2^2+\gamma_3^2\big)(e_1+e_2+e_3)-e_1\gamma_1^2+e_2\gamma_2^2+e_3\gamma_3^2, \\
\rho = e_2e_3\gamma_1^2+e_1e_3\gamma_2^2+e_1e_2\gamma_3^2
\end{gather*}
are the same polynomials of second order in variables $\gamma$ as above~\eqref{sigma-rho}.
\end{proposition}

Following to S.A.~Chaplygin \cite{ch03} we can introduce the sphero-conical coordinates $u_1$, $u_2$
 \begin{gather*}
\gamma_i=\sqrt{\frac{(u_1-e_i)(u_2-e_i)}{(e_j-e_i)(e_m-e_i)}}
,\qquad i\neq j\neq m,
\end{gather*}
and explicitly present some solutions of the Bertrand--Darboux type equations (\ref{uv-eqch}). Namely, integrals of motion~$H_{1,2}$~(\ref{ch-ham}) satisfy the separation relations
\begin{gather*}
\frac{4(e_1-u_i)(e_2-u_i)(e_3-u_i)}{d(e_1+1)(e_2+1)(e_3+1)}p_{u_i}^2+G_i(u_i)-\frac{u_iH_2}{d(u_i+1)}+2H_1=0,\qquad i=1,2,
\end{gather*}
which can be considered as gentle deformation of the initial relations (\ref{ell-sep-rel}). Thus, separable potentials in this case read as
\begin{gather*}
U_2 = \frac{d(u_1+1)(u_2+1) (G_1(u_1)-G_2(u_2) )}{u_2-u_1}, \\
U_1 = \frac{u_1u_2 (G_1(u_1)-G_2(u_2) )+u_2G_1(u_1)-u_1G_2(u_2)}{2(u_2-u_1)}.
\end{gather*}
The passage to limit $d\to 0$ reduces equations of motion for the Chaplygin ball~(\ref{ch-eqV}) to the standard Euler--Poisson equations. However, at $d\to 0$ we have to simultaneously change the def\/inition of the second potential~$U_2$ in~(\ref{zam-ch}) and, therefore, we present another family of solutions for equations~(\ref{uv-eqch}).

Let us introduce variables $v_{1,2}$
 \begin{gather*}
\gamma_i=\sqrt{\frac{(1-da_j)(1-da_m)}{(1-dv_1)(1-dv_2)}}
 \cdot \sqrt{\frac{(v_1-a_i)(v_2-a_i)}{(a_j-a_i)(a_m-a_i)}} ,\qquad i\neq j\neq m,
\end{gather*}
and the conjugated momenta $p_{v_{1,2}}$, see~\cite{ts11} for details. In this variables the separated relations have the following form
\begin{gather*}
 4(1-dv_i)(v_i-a_1)(v_i-a_2)(v_i-a_3)p_{v_i}^2+U_i(v_i)+v_iH_2-2H_1=0, \qquad i=1,2,
\end{gather*}
and integrable potentials
\begin{gather*}
U_2=\frac{G_1(v_1)-G_2(v_2)}{v_2-v_1},\qquad U_1=\frac{v_2G_1(v_1)-v_1G_2(v_2)}{2(v_2-v_1)}
\end{gather*}
are the same functions on variables $v_{1,2}$ as the integrable potentials on the sphere (\ref{uv-sol}).

\begin{proposition}
At $C_2=0$ vector field \eqref{ch-eqV} has integrals of motion of second order in veloci\-ties~\eqref{ch-ham} if potentials~$U_{1,2}$
have the same form as integrable potentials on the sphere~\eqref{uv-nell}
\begin{gather*}
U_2^{(N)} = \sum_{k=0}^{[N/2]}(-1)^k\left(
 \begin{matrix}
 N-k \\
 k
 \end{matrix}
 \right) \varrho^k\varsigma^{N-2k}, \\
 2U_1^{(N)} = \sum_{k=0}^{[(N-1)/2]}(-1)^k\left(
 \begin{matrix}
 N-k -1\\
 k
 \end{matrix}
 \right) \varrho^{k+1}\varsigma^{N-2k-1},
\end{gather*}
and
\begin{gather*}
U_2^{(-K)}=-\frac{U_1^{(K)}}{\varrho^K},
\qquad
2 U_1^{(-K)}
=-\frac{U_1^{(K+1)}}{\varrho^K}.
\end{gather*}
Of course, any linear combination of these polynomial and rational potentials also satisfies the equations~\eqref{uv-eqch}.
These potentials differ from~\eqref{uv-nell} by replacement of polynomials $\sigma$ and $\rho$ for the following functions
\begin{gather*}
\varsigma = \mathrm g(\gamma)\bigl(\sigma+d\bigl(a_1(a_2+a_3)\gamma_1^2+a_2(a_1+a_3)\gamma_2^2+a_3(a_1+a_2)\gamma_3^2\bigr)\bigr), \\
\varrho = \mathrm g(\gamma)\bigl(\rho+da_1a_2a_3 \big(\gamma_1^2+\gamma_2^2+\gamma_3^2\big)\bigr),
\end{gather*}
which at $d=0$ become initial polynomials~\eqref{sigma-rho}.
\end{proposition}

For instance, at $N=2$ we have the following analogue of the Neumann system
\begin{gather*}
H_1^{(2)}=\frac{1}{2} (M,\mathbf A M)+\varrho,\qquad H_2^{(2)}=(M,M)+\varsigma,
\end{gather*}
 and at $K=1$ the following counterpart of the Braden system
 \begin{gather*}
H_1^{(-1)}=\frac{1}{2} (M,\mathbf A M)-\frac{\varsigma}{\varrho},\qquad
H_2^{(-1)}=(M,M)-\frac{1}{\varrho}.
\nonumber
\end{gather*}
Of course, we can also single out other families of solutions of the equations~(\ref{uv-eqch}), for instance, see~\cite{dr98}.

\section{Nonholonomic oscillator and Heisenberg system}\label{section6}

Let us consider the Lagrangian of the particle in Euclidean space~$\mathbb R^3$
\begin{gather}\label{3-lag}
L=\frac{m}{2}\big(\dot{x}^2+\dot{y}^2+\dot{z}^2\big)-V(x,y,z),
\end{gather}
where $m$ is the mass of the particle. When this system is subject to a nonholonomic constraint, the resulting mechanical system may or may not preserve energy and the phase space volume, and their integrability and
reduction theories are completely dif\/ferent from the Hamiltonian case~\cite{bcus99,koiller04, guha13}. In this Section we consider two f\/irst-order nonholonomic constraints, which displays all the basic properties of f\/irst-order nonholonomic systems in the control theory~\cite{bl03}.

The f\/irst nonholomic constraint and potential in~(\ref{3-lag}) for the so-called nonholonomic oscillator have the following form
\begin{gather*}
f=\dot{z}-ky\dot{x}=0,\qquad V=\frac{y^2}{2}\qquad k\in\mathbb R,
\end{gather*}
whereas second constraint and potential in (\ref{3-lag}) for the so-called Heisenberg system read as
\begin{gather*}
f=\dot{z}-(y\dot{x}-x\dot{y})=0,\qquad V=0.
\end{gather*}
The Heisenberg system (nonholonomic integrator) can be pointed out as a benchmark example
of nonholonomic system with a first-order nonintegrable constraint, which mimics the kinematic model of a wheeled mobile
robot of the unicycle type.

In generic case at $V(x,y,z)=V(x,y)$ in (\ref{3-lag}) the third degree of freedom also decouples from the rest of the system and after nonholonomic reduction we obtain a two-degrees of freedom system of the Chaplygin type \cite{bl03,guha13,mol12}.
For the such generalized nonholonomic oscillator the reduced equations of motion are
\begin{gather}\label{red-osc}
\dot{x}=\frac{p_x}{m},\qquad \dot{y}=\frac{p_y}{m},\qquad
\dot{p}_x=-\frac{1}{1+k^2y^2}\bigl(k^2yp_xp_y+\partial_x V\bigr),\qquad
\dot{p}_y=-\partial_y V.
\end{gather}
For the generalized Heisenberg system the reduced equations of motion read as
\begin{gather}
\dot{x}=\frac{p_x}{m},\qquad \dot{y}=\frac{p_y}{m},\nonumber\\
\dot{p}_x=-\frac{(x^2+1)\partial_xV+xy\partial_yV}{m(1+x^2+y^2)},\qquad
\dot{p}_y=-\frac{(y^2+1)\partial_yV+xy\partial_xV}{m(1+x^2+y^2)}.\label{red-hs}
\end{gather}
Below we will study potentials $V(x,y)$ in~(\ref{red-osc}) and~(\ref{red-hs}), so that the corresponding four-dimensional vector f\/ields~$X$ have an additional f\/irst integral and possess an invariant volume form. Of course, equations on these potentials have the form of the characteristic equation~(\ref{char-eq}) and can be considered as an analogue of the Bertrand--Darboux equation~(\ref{bd-eq}).

\subsection{The generalised nonholonomic oscillator}
The vector f\/ield for the reduced nonholonomic oscillator (\ref{red-osc})
after the following change of variables
\begin{gather*}
p_1 =\sqrt{k^2y^2+1} p_x,\qquad p_2 = \frac{p_y}{\sqrt{k^2y^2+1}},\qquad q_1=x,\qquad q_2=y
\end{gather*}
becomes the conformally Hamiltonian vector f\/ield
\begin{gather*}
X=-\mu P{\mathrm d}H_1,\qquad P=\left(
 \begin{matrix}
 0 & \mathrm I \\
 -\mathrm I & 0
 \end{matrix}
 \right)
\end{gather*}
with respect to the canonical Poisson bivector $P$ and reduced Hamiltonian
\begin{gather*}
H_1=\sum_{i,j=1}^2\mathrm g_{ij}p_ip_j+V(q_1,q_2)=\frac{p_1^2}{2m}+\frac{p_2^2(k^2q_2^2+1)}{2m}+V(q_1,q_2).
\end{gather*}
Conformal factor
\begin{gather*}
\mu=\frac{1}{\sqrt{k^2q_2^2+1}}
\end{gather*}
 is a nowhere vanishing smooth function on an open dense subset of the plane $q_2\neq \infty$, which def\/ines an invariant volume form $\hat{\Omega}=\mu dq\wedge dp$.

Substituting a linear function in velocities
\begin{gather*}
H_2=g_1(q_1,q_2)p_1+g_2(q_1,q_2)p_2
\end{gather*}
into the equation $\dot{H}_2=0$ one gets
\begin{gather*}
g_1 =c_1+c_2\ln\bigl(kq_2+\sqrt{k^2q_2^2+1}\bigr),\qquad
g_2 = -\sqrt{k^2q_2^2+1} (c_2kq_1-c_3)
\end{gather*}
and
\begin{gather*}V(q_1,q_2)=G\left(-\frac{c_1\ln\bigl(kq_2+\sqrt{k^2q_2^2+1}\bigr)}{k}-c_2\left(\frac{kq_1^2}{2}
+\int\! \frac{ \ln\bigl(kq_2+\sqrt{k^2q_2^2+1}\bigr)}{\sqrt{k^2q_2^2+1}}\right)\!-c_3q_1\right).
\end{gather*}
If we want to consider a single valued integral $H_2$, we have to put $c_2=0$ and
\begin{gather*}
c_3=0,\qquad V=G(q_1)\qquad \mbox{or}\qquad c_1=0,\qquad V=G(q_2).
\end{gather*}
Substituting polynomials of second order in velocities
\begin{gather*}
H_2=\sum_{i,j=1}^2 K_{ij}(q_1,q_2) p_ip_j+U(q_1,q_2),
\end{gather*}
where $K_{ij}$ and $U$ are single valued functions on $q_1$, $q_2$ in the equation $\dot{H}_2=0$, we obtain the following expression for the second integral of motion
\begin{gather*}
H_2=\frac{c_1\big(k^2q_2^2+kq_2\sqrt{k^2q_2^2+1}+1\big)}{kq_2+\sqrt{k^2q_2^2+1}}p_1p_2+c_2\big(k^2q_2^2+1\big)p_2^2+U(q_1,q_2),
\end{gather*}
and the following counterpart of the Bertrand--Darboux equation
\begin{eqnarray}
\frac{c_1}{\sqrt{k^2q_2^2+1}}\bigl(\big(k^2q_2^2+1\big)\partial_{22} V+k^2q_2\partial_2V-\partial_{11}V\bigr)
-2 c_2\partial_{12} V=0.
\end{eqnarray}
This equation has one physical and one formal solution
\begin{gather*}
c_1=0,\qquad V=G_1(q_1)+G_2(q_2)\qquad\mbox{and}\qquad
c_2=0,\qquad V=G_1(q_+)+G_2(q_-),
\end{gather*}
where
\begin{gather*}
q_\pm=q_1\pm\frac{\ln\bigl(kq_2+\sqrt{k^2q_2^2+1} \bigr)}{k}.
\end{gather*}
Thus, for the nonholonomic oscillator we obtain only trivial perturbations in the framework of the Bertrand--Darboux method.

\subsection{The generalized Heisenberg system}
The vector f\/ield for the reduced Heisenberg system (\ref{red-hs})
after the following change of variables
\begin{gather*}
p_1 =\frac{m \bigl((1+y^2)p_x-xyp_y\bigr)}{1+x^2+y^2},\qquad p_2 =\frac{m \bigl((1+x^2)p_y-xyp_x\bigr)}{1+x^2+y^2},\qquad q_1=x,\qquad q_2=y
\end{gather*}
is conformally Hamiltonian vector f\/ield
\begin{gather}\label{conf-ham}
X=-\mu P{\mathrm d}H_1,
\qquad P=\left(
 \begin{matrix}
 0 & \mathrm I \\
 -\mathrm I & 0
 \end{matrix}
 \right)
\end{gather}
with respect to canonical Poisson bivector $P$ and reduced Hamiltonian
\begin{gather*}
H_1=\sum_{i,j=1}^2 \mathrm g_{ij}p_ip_j+V(q_1,q_2)=\frac{q_1^2+q_2^2+1}{2m}\bigl(p_1^2+p_2^2+(q_1p_1+q_2p_2)^2\bigr)+V(q_1, q_2).
\end{gather*}
Conformal factor
\begin{gather*}
\mu=\big(1+q_1^2+q_2^2\big)^{-1}
\end{gather*}
is a nowhere vanishing smooth function on an open dense subset of the plane $q_{1,2}\neq \infty$, which def\/ines an invariant volume form $\hat{\Omega}=\mu dq\wedge dp$.

Substituting linear function in velocities
\begin{gather*}
H_2=g_1(q_1,q_2)p_1+g_2(q_1,q_2)p_2
\end{gather*}
into the equation $\dot{H}_2=0$ one gets the following f\/irst integral
\begin{gather*}
H_2=\big(p_1 q_1^2+p_2 q_1 q_2+p_1\big) c_1+
\big(p_1 q_1 q_2+p_2 q_2^2+p_2\big) c_2+
(p_1 q_2-p_2 q_1) c_3,\qquad c_k\in\mathbb R,
\end{gather*}
and potential
\begin{gather*}
V=G\left(\frac{(c_1 q_1+ c_2q_2)^2+2 c_3(c_1 q_2- c_2 q_1)+c_1^2+c_2^2-c_3^2}{ (c_1 q_2-c_2 q_1-c_3)^2}\right)
\end{gather*}
depending on the arbitrary function $G$.

Substituting polynomials of second order in velocities
\begin{gather*}
H_2=\sum_{i,j=1}^2 K_{ij}(q_1,q_2) p_ip_j+U(q_1,q_2)
\end{gather*}
 into the equation $\dot{H}_2=0$ one gets the following expression for the second integral of motion
\begin{gather*}
H_2 = (p_1 q_2-p_2 q_1) (p_1 q_1^2+p_2 q_1 q_2+p_1) c_1+\big(p_1 q_1^2+p_2 q_1 q_2+p_1\big) \big(p_1 q_1 q_2+p_2 q_2^2+p_2\big) c_2 \\
\hphantom{H_2}{}
+ (p_1 q_2-p_2 q_1) \big(p_1 q_1 q_2+p_2 q_2^2+p_2\big) c_3-(p_1 q_2-p_2 q_1)^2c_4 \\
\hphantom{H_2}{} + \bigl(q_2^2\big(q_1^2+1\big)p_1^2+2q_2^3q_1p_1p_2+\big(q_2^4+q_1^2+2q_2^2+1\big)p_2^2\bigr) c_5 \\
\hphantom{H_2}{} + \bigl(\big(q_1^4+2 q_1^2+q_2^2+1\big) p_1^2+2 q_1^3 q_2 p_1 p_2+q_1^2 \big(q_2^2+1\big) p_2^2\bigr) c_6+U(q_1,q_2).
\end{gather*}
In this case equation (\ref{char-eq}) looks like
\begin{gather*}
A\partial_{11}V+2B\partial_{12}V+C\partial_{22}V + \frac{1}{1+q_1^2+q_2^2}\bigl(a\partial_1V+b\partial_2V\bigr)=0,
\end{gather*}
where $A$, $B$, $C$ are the polynomials of second order in $q_{1,2}$
\begin{gather*}
A = \big(q_1^2+1\big)(q_1c_1-c_2)+q_2\big(q_1^2-1\big)c_3+2q_1q_2(c_6-c_4), \\
B = q_2\big(q_1^2+1\big)c_1+q_1\big(q_2^2+1\big)c_3+\big(q_1^2-q_2^2\big)c_4-\big(q_1^2+1\big)c_5+\big(q_2^2+1\big)c_6,\\
C = \big(q_2^2+1\big)(q_2c_3+c_2)+q_1\big(q_2^2-1\big)c_1+2q_1q_2(c_4-c_5),
\end{gather*}
and $a$, $b$ are the polynomials of fourth order
\begin{gather*}
a = \big(2 q_1^4+2 q_1^2 q_2^2+5 q_1^2-q_2^2+3\big) c_1+q_1 \big(q_1^2-3 q_2^2+1\big) c_2+2 q_1 q_2 \big(q_1^2+q_2^2+3\big) c_3 \\
\hphantom{a =}{} - 2 q_2 \big(q_1^2+q_2^2+3\big) c_4+4 q_2 \big(q_1^2+1\big) c_5-2 q_2 \big(q_1^2-q_2^2-1\big) c_6, \\
b = 2 q_1 q_2 \big(q_1^2+q_2^2+3\big) c_1+q_2 \big(3 q_1^2-q_2^2-1\big) c_2+\big(2 q_1^2 q_2^2+2 q_2^4-q_1^2+5 q_2^2+3\big) c_3 \\
\hphantom{b =}{} + 2 q_1 \big(q_1^2+q_2^2+3\big) c_4-2 q_1 \big(q_1^2-q_2^2+1\big) c_5-4 q_1 \big(q_2^2+1\big) c_6.
\end{gather*}
Following Darboux~\cite{darb} we can f\/ind the canonical form of the corresponding Killing tensor and a few families of solutions to this equation. For instance, if $c_6=1$ and other constants of integration are equal to zero, then solutions of the equation~$\dot{H}_2=0$ are labelled by two arbitrary functions~$G_{1,2}$
\begin{gather*}
V(q_1,q_2) = \frac{q_1^2+q_2^2+1}{2m} G_1(q_2)+\frac{1}{2m} G_2\left(\frac{q_2^2+1}{q_1^2}\right), \\
U(q_1,q_2)=q_1^2G_1(q_2)+G_2\left(\frac{q_2^2+1}{q_1^2}\right).
\end{gather*}
If $c_4=1$ and other constants of integration are equal to zero, then we have solution
\begin{gather*}
 V(q_1, q_2) =G_1(r) -\frac{r^2+1}{2mr^2} G_2(\varphi),\qquad
 U(q_1, q_2) = G_2(\varphi)
\end{gather*}
associated with polar coordinates on the plane
\begin{gather*}
q_1=r\cos\varphi,\qquad q_2=r\sin\varphi.
\end{gather*}
Here $G_{1,2}$ are arbitrary functions. In similar manner we can get solutions associated with parabolic and elliptic coordinates on the plane, but they are bulky and, therefore, we do not present these solution explicitly.

In order to get these solutions we can also use the Birkhof\/f method. Namely, let us consider a general natural system of two degrees of freedom described in certain generalized coordinates by the following Lagrangian
\begin{gather*}
L=\sum_{i,j=1}^{2}\mathrm g_{ij}(q)\frac{dq_i}{dt}\frac{dq_j}{dt}-V(q_1,q_2).
\end{gather*}
According to Birkhof\/f~\cite{bir27} using change of time $t\to \tau$ and
coordinates $(q_1,q_2) \to (x,y)$ this Lagrangian can always be reduced to the form of
\begin{gather*}
L=\left(\frac{dx}{d\tau}\right)^2+\left(\frac{dy}{d\tau}\right)^2-U(x,y) .
\end{gather*}
Thus, taking the well-known solutions $U(x,y)$ of the classical Bertrand--Darboux equation~(\ref{bd-eq}) and applying the inverse Birkhof\/f transformation we are able to obtain integrable potentials $V(q_1,q_2)$ for the two-dimensional holonomic system with nonstandard metric.

We can apply this method for the given nonholonomic case because the corresponding vector f\/ield~$X$ (\ref{conf-ham}) is a conformally Hamiltonian vector f\/ield~\cite{mol12,ts12}, i.e., it can be reduced to a~Hamiltonian vector f\/ield by changing of time. Recall, that equation $\dot{H}_2=0$~(\ref{m-eqint}) is invariant with respect to change of time, which we have to use both in the Birkhof\/f method and in the reduction of the conformally Hamiltonian vector f\/ield to the Hamiltonian one.

\section{Conclusion}\label{section7}

In this paper, we consider perturbations of the f\/ive well-known two-dimensional nonholonomic systems, which are integrable by the Euler--Jacobi theorem. We show that the Bertrand--Darboux method is applicable to these systems and that all the obtained Bertrand--Darboux type equations in nonholonomic case can be reduced to the Bertrand--Darboux type equations in holonomic case.
Consequently, we can directly obtain all the possible integrable potentials for these nonholonomic systems directly from the well-known integrable potentials of the Hamiltonian mechanics.

\subsection*{Acknowledgements}
We are greatly indebted B.~Jovanovi\'{c} and the anonymous referees for a~relevant contribution to improve the paper.
The work on the revised, f\/inal version of this paper was supported by Russian Science Foundation (project No~15-12-20035).

\pdfbookmark[1]{References}{ref}
\LastPageEnding

\end{document}